\documentclass[11pt]{article}
\usepackage{graphicx}

\begin{document}
%


\begin{center}
{\large \bf What happened to the Cosmological QCD Phase
Transition?} \vskip.5cm

W-Y. P. Hwang\footnote{Email: wyhwang@phys.ntu.edu.tw} \\
{\em The Research Center for Cosmology and Particle Astrophysics,
\\Center for Theoretical Sciences, Institute of Astrophysics,
\\and Department of Physics, National
Taiwan University, Taipei 106, Taiwan} \vskip.2cm


{\small(Originally Written in October 22, 2006; Revised in April
17, 2007)}
\end{center}

\begin{abstract}
The scenario that some first-order phase transitions may have
taken place in the early Universe offers us one of the most
intriguing and fascinating questions in cosmology. Indeed, the
role played by the latent "heat" or energy released in the phase
transition is highly nontrivial and may lead to some surprising,
important results. In this paper, we take the wisdom that the
cosmological QCD phase transition, which happened at a time
between $10^{-5}\, sec$ and $10^{-4}\,sec$ or at the temperature
of about $150\, MeV$ and accounts for confinement of quarks and
gluons to within hadrons, would be of first order. To get the
essence out of the scenario, it is sufficient to approximate the
true QCD vacuum as one of degenerate $\theta$-vacua and when
necessary we try to model it effectively via a complex scalar
field with spontaneous symmetry breaking. We examine how and when
"pasted" or "patched" domain walls are formed, how long such walls
evolve in the long run, and we believe that the significant
portion of dark matter could be accounted for in terms of such
domain-wall structure and its remnants. Of course, the
cosmological QCD phase transition happened in the way such that
the false vacua associated with baryons and many other
color-singlet objects did not disappear (that is, using the
bag-model language, there are bags of radius 1.0 fermi for the
baryons) - but the amount of the energy remained in the false
vacua is negligible. The latent energy released due to the
conversion of the false vacua to the true vacua, in the form of
"pasted" or "patched" domain walls in the short run and their
numerous evolved objects, should make the concept of the
"radiation-dominated" epoch, or of the "matter-dominated" epoch to
be re-examined.

\medskip

{\parindent=0truept PACS Indices: 12.39.Ba, 12.38.Mh, 12.38.Lg,
98.80.Bp, 98.80.-k.}
\end{abstract}

\section{Introduction}
The discovery \cite{DMR} of fluctuations or anisotropies, at the
level of $10^{-5}$, associated with the cosmic microwave
background (CMB) has helped transformed the physics of the early
universe into a main-stream research area in astronomy and in
particle astrophysics, both theoretically and observationally
\cite{Bennett}. CMB anisotropies\cite{WMAP1} and
polarizations\cite{WMAP3}, the latter even smaller and at the
level of $10^{-7}$, either primary (as imprinted on the last
scattering surface just before the universe was $(379\pm 8)\times
10^3$ years old) or secondary (as might be caused by the
interactions of CMB photons with large-scale structures along the
line of sight), are linked closely to the inhomogeneities produced
in the early universe.

Over the last three decades, the standard model of particle
physics has been well established to the precision level of
$10^{-5}$ or better in the electroweak sector, or to the level of
$10^{-3}-10^{-2}$ for the strong interactions. In the theory, the
electroweak (EW) phase transition, which endows masses to the
various particles, and the QCD phase transition, which gives rise
to confinement of quarks and gluons within hadrons in the true QCD
vacuum, are two well-established phenomena. Presumably, the EW and
QCD phase transitions would have taken place in the early
universe, respectively, at around $10^{-11}\, sec$ and at a time
between $10^{-5}\,sec$ and $10^{-4}\,sec$, or at the temperature
of about $300\,GeV$ and of about $150\, MeV$, respectively.
Indeed, it has become imperative to formulate the EW and QCD phase
transitions in the early universe if a quantitative theory of
cosmology can ever be reached.

The purpose of this work is to focus our attention on cosmological
QCD phase transition and to assess whether its roles in the early
universe can be synthesized in a more quantitative terms - a
problem which has become one of the most challenging problems in
the physics of the early universe. To simplify the situation, we
use the bag-model language and try to model the degenerate
$\theta$-vacua, the lower-temperature phase, as the minima of the
spontaneously-broken complex scalar fields. In particular, we try
to set up the strategy of tackling the problem by dividing it into
problems in four different categories, viz.: (1) how a bubble of
different vacuum grows or shrinks; (2) how two growing bubbles
collide or squeeze, and merging, with each other; (3) how the
Universe eventually stabilize itself later while keeping expanding
for several orders of magnitude; and (4) how specific objects,
such as back holes or magnetic strings, get produced during the
specific phase transition. Questions related to part (4), which
are quite complicated, will not be addressed here; see, e.g., ref.
\cite{Zhitnitsky}. In the framework which we consider, we could
describe the intermediate solutions based on the so-called
"pasted" or "patched" domain walls when the majority of the false
vacua get first eliminated - but how it would evolve from there
and how long it would evolve still uncertain.

{\bf The major result of this paper is that the latent heat (or
latent energy), which turns out to be identified as the "bag
constant", is huge compared to the radiation density at the
cosmological QCD phase transition (i.e. at about $3\times 10^{-5}
sec$). As time evolved to the present, the percentage of this
quantity becomes probably the majority of dark matter (25 \% of
the composition of the present Universe).}

\section{The Background Universe as from Einstein's General Relativity
 and the Cosmological Principle}

A prevailing view regarding our universe is that it originates
from the joint making of Einstein's general relativity and the
cosmological principle while the observed anisotropies associated
with the cosmic microwave background (CMB), at the level of about
one part in 100,000, may stem, e.g., from quantum fluctuations in
the inflation era. In what follows, we wish to first outline very
briefly a few key points in the standard scenario so that we shall
have a framework which we may employ to elucidate the roles of
phase transitions in the early universe.

Based upon the cosmological principle which state that our
universe is homogeneous and isotropic, we use the Robertson-Walker
metric to describe our universe\cite{Turner}.
\begin{equation}
ds^2=dt^2 -R^2(t)\{ {dr^2\over 1-kr^2} +r^2 d\theta^2
+r^2 sin^2\theta d\phi^2\}.
\end{equation}
Here the parameter $k$ describes the spatial curvature with
$k=+1$, $-1$, and $0$ referring to an open, closed, and flat
universe, respectively. The scale factor $R(t)$ describes the size
of the universe at time $t$.

To a reasonable first approximation, the universe can be described
by a perfect fluid, i.e., a fluid with the energy-momentum tensor
$T^\mu \ _\nu = \, diag\, (\rho,\ ,\ -p,\ -p,\, -p)$ where $\rho$
is the energy density and $p$ the pressure. Thus, the Einstein
equation, $G^\mu\ _\nu = 8\pi G_N T^\mu\ _\nu + \Lambda g^\mu\
_\nu$, gives rise to only two independent equations, i.e., from
$(\mu,\ \nu) = (0,\ 0)$ and $(i,\ i)$ components,

\begin{equation}
{{\dot R}^2\over R^2}+{k\over R^2} = {8\pi G_N\over 3}\rho +{\Lambda \over 3}.
\end{equation}

\begin{equation}
2{\ddot R\over R}+ {{\dot R}^2\over R^2}+{k\over R^2} = -8\pi G_N\  p +\Lambda.
\end{equation}
Combining with the equation of state (EOS), i.e. the relation
between the pressure $p$ and the energy density $\rho$, we can
solve the three functions $R(t)$, $\rho(t)$, and $p(t)$ from the
three equations.  Further, the above two equations yields
\begin{equation}
{\ddot R\over R} = -{4\pi G_N\over 3} (\rho +3 p) + {\Lambda \over 3},
\end{equation}
showing either that there is a positive cosmological constant or
that $\rho + 3p$ must be somehow negative, if the major conclusion
of the Supernovae Cosmology Project is correct \cite{SNIa}, i.e.
the expansion of our universe still accelerating (${\ddot R/ R}
>0$).

Assuming a simple equation of state, $p= w \rho$, we obtain, from Eqs. (2) and (3),

\begin{equation}
2 {\ddot R\over R}+ (1+ 3w)({\dot R^2\over R^2} + {k\over R^2})- (1+w)\Lambda=0,
\end{equation}
which is applicable when a particular component dominates over the
others - such as in the inflation era (before the hot big bang
era), the radiation-dominated universe (e.g. the early stage of
the hot big bang era), and the matter-dominated universe (i.e.,
the late stage of the hot big bang era, before the dark energy
sets in to dominate everything else). {\it In light of
cosmological QCD phase transition, we would like to examine if the
radiation-dominate universe and the matter-dominated universe
could ever exist at all, since this has become a dogma in the
thinking of our Universe.}

For the {\bf Inflation Era}, we could write $p= -\rho$ and $k=0$
(for simplicity), so that
\begin{equation}
\ddot R - {\dot R^2\over R}=0,
\end{equation}
which has an exponentially growing, or decaying, solution $R
\propto e^{\pm \alpha t}$, compatible with the so-called
"inflation" or "big inflation". In fact, considering the simplest
case of a real scalar field $\phi(t)$, we have
\begin{equation}
\rho = {1\over 2} {\dot \phi}^2 + V(\phi), \qquad p ={1\over 2} {\dot \phi}^2 -V(\phi),
\end{equation}
so that, when the "kinetic" term ${1\over 2}{\dot \phi}^2$ is
negligible, we have an equation of state, $p \sim -\rho$. In
addition to its possible role as the "inflaton" responsible for
inflation, such field has also been invoked to explain the
accelerating expansion of the present universe, as dubbed as
"quintessence" or "complex quintessence"\cite{Quint}.

Let's look at the standard textbook argument leading to the
radiation-dominated universe and the matter-dominated universe:

For the {\bf Radiation-Dominated Universe}, we have $p=\rho/3$.
For simplicity, we assume that the curvature is zero ($k=0$) and
that the cosmological constant is negligible ($\Lambda =0$). In
this case, we find from Eq. (5)
\begin{equation}
R \propto t^{1\over 2}.
\end{equation}

Another simple consequence of the homogeneous model is to derive
the continuity equation from Eqs. (2) and (3):
\begin{equation}
d (\rho R^3) +p d(R^3)=0.
\end{equation}
Accordingly, we have $\rho\propto R^{-4}$ for a
radiation-dominated universe ($p = \rho/3$) while $\rho \propto
R^{-3}$ for a matter-dominated universe ($p << \rho$).  The
present universe is believed to have a matter content of about
5\%, or of the density of about $5 \times 10^{-31} g/cm^3$, much
bigger than its radiation content $5 \times 10^{-35} g/cm^3$, as
estimated from the $3^\circ$ black-body radiation. However, as $t
\to 0$, we anticipate $R \to 0$, extrapolated back to a very small
universe as compared to the present one. Therefore, the universe
is necessarily dominated by the radiation during its early enough
epochs.

For the radiation-dominated early epochs of the universe with
$k=0$ and $\Lambda =0$, we could deduce, also from Eqs. (2) and
(3),
\begin{equation}
\rho ={3\over 32 \pi G_N}t^{-2},\qquad T =\{ {3c^2\over 32 \pi G_N a}\}^{1\over 4}
t^{-{1\over 2}} \cong 10^{10}t^{-1/2} (^\circ K).
\end{equation}
These equations tell us a few important times in the early universe, such as $10^{-11} sec$
when the temperature $T$ is around $300\ GeV$ during which the electroweak (EW) phase
transition is expected to occur, or somewhere between $10^{-5} sec$ ($\cong 300\ MeV$)
and $10^{-4}sec$ ($\cong 100\ MeV$) during which quarks and gluons undergo the QCD
confinement phase transition.

For the {\bf Matter-Dominated Universe}, we have $p\approx 0$,
together with the assumption that $k=0$ and $\Lambda=0$. Eq. (5)
yields
\begin{equation}
R \propto t^{2\over 3}.
\end{equation}
As mentioned earlier, the matter density $\rho_m$ scales like
$R^{-3}$, or $\rho_m \propto t^{-2}$, the latter similar in the
radiation-dominated case.

When $t=10^9 sec$, we have $\rho_\gamma=6.4\times 10^{-18}gm/cm^3$
and $\rho_m=3.2\times 10^{-18}gm/cm^3$, which are close to each
other and it is almost near the end of the radiation-dominated
universe. The present age of the Universe is 13.7 billion years -
for a large part of it, it is matter-dominated although now we
have plenty of dark energy (65\% $\sim$ 70\%).

However, it is generally believed that our present universe is
already dominated by the dark energy (the simplest form being of
the cosmological constant; about 70\%) and the dark matter (about
25\%). The question is when this was so - when the dark part
became dominant.

There is another basic point - trivial but important. For both the
electroweak and QCD phase transitions in the early Universe, if
the phase transitions are described (approximately) by the complex
fields $\phi$, then the density of the system is given by
\begin{equation}
\rho=\rho_\phi+\rho_\gamma+\rho_m+...,
\end{equation}
before or after or during the phase transition is being taking
place. For the electroweak or QCD phase transition, we know that
$\rho_m << \rho_\gamma$, but the role played by $\rho_\phi$ is
clearly of importance in our considerations.

{\bf What would be missing in the standard textbook argument?} We
would come back in Section 6 to this important point, after we set
up the general framework and have gained enough of insights. The
crucial point is whether cosmological QCD phase transition is the
first-order phase transition - {\it if it is, there is the latent
"heat" or energy released in the transition; the story would
change dramatically if the amount of energy density turns out to
be greater than either $\rho_\gamma$ or $\rho_m$ in the previous
radiation-dominated or matter-dominated era.} We have to question
what happened if $\rho_\gamma$ would no longer be dominant in the
radiation-dominated universe - the "new" dominant sector of the
universe may not influence the "old" radiation-dominated piece but
Einstein equation in principle no longer guarantee its validity
(about the old, no-longer-dominant sector).


\section{The Cosmological QCD Phase Transition - the Big Picture}

Let's try to focus on the QCD phase transition in the early
Universe, or on the cosmological QCD phase transition.

At the temperature $T>T_c\sim 150 MeV$, i.e., before the phase
transition takes place, free quarks and gluons can roam anywhere.
As the Universe expands and cools, eventually passing the critical
temperature $T_c$, the bubbles nucleate here and there. These
bubbles "explode", as we call it "exploding solitons". When it
reaches the "supercooling" temperature, $T_s$, or something
similar, the previous bubbles become too many and in fact most of
them become touched each other - now the false vacua or "bubbles"
of different kind (where quarks and gluons can move freely) start
to collapse - or we call it "imploding solitons". When all these
bubbles of different kind implode completely, the phase transition
is now complete.

There is some specialty regarding the QCD phase transition in the
early Universe. Namely, the collapse of the false vacuum does
depend on the inside quark-gluon content - e.g., if we have a
three-quark color-singlet combination inside, the collapse of the
false vacuum would stop (or stabilize) at a certain radius (we
called the bag radius, like in the MIT bag radius); of course,
there are meson configurations, glueballs, hybrids, six-quark or
multi-quark configurations, etc. The QCD phase transition in the
early Universe does not eliminate all the false vacua; rather, the
end state of the transition could have at least lots of baryon or
meson states, each of them has some false vacuum to stabilize the
system.

How big can a bubble grow? It is with the fastest speed which the
bubble can grow is through the speed of light or close to the
speed of light. The bubble could sustain from the moment it
creates, say, $T\approx T_c$ to the moment of supercooling,
$T_s\sim 0.95 \cdot T_c $, or during the time span $t \sim 3\times
10^{-5} \times 0.05 sec $ (or $1.5 \times 10^{-7} sec$). So, the
bubble can at most grow into $c \cdot 1.5 \times 10^{-7} sec$ or
$4.5 \times 10^{3}\, cm$.

How big was the Universe during the cosmological QCD phase
transition? Compared to the size now, an estimate is the expansion
of $5.7\times 10^{12}$, a huge factor. (See the beginning of Sect.
6.) In the simplest approximation (when the scalar fields don't
couple to the other fields, such as gluons or quarks), the domain
walls cannot disappear - not only sometime because of the possible
nontrivial topology but that there should be some QCD dynamics to
annihilate the walls. In light of the huge expansion factor, the
domain wall structure cannot survive, except the strict topology
(which we call "domain-wall nuggets").

As a yardstick, we note that, at $t \sim 10^{-5}\,sec$ or $T\sim
300\,MeV$, we have
\begin{equation}
\rho_\gamma\sim 6.4 \times 10^{10}gm/cm^3,\qquad \rho_m\sim
3.2\times 10^3 gm/cm^3.
\end{equation}
Or, at $t\sim 3.30 \times 10^{-5}\,sec$ or $T=T_c\sim 150 MeV$, we
have
\begin{equation}
\rho_\gamma= 5.88\times 10^9 gm/cm^3, \qquad \rho_m =6.51\times
10^2 gm/cm^3.
\end{equation}
Slightly later when QCD phase transition has completed, at $t\sim
10^{-4}\,sec$ or $T\sim 100\, MeV$, we have
\begin{equation}
\rho_\gamma\sim 6.4 \times 10^8 gm/cm^3,\qquad \rho_m \sim 1.0
\times 10^2 gm/cm^3.
\end{equation}

In what follows, we use the so-called "bag models"\cite{MIT,TDLee}
to have the simplified version of quark confinement - I think it
is important to use the simplified version in the complicated
cosmological environment, in order to extract meaningful results.

When the low-temperature bubbles start to show up (i.e. to
nucleate), it is about $T_c\approx T < T_s$. This period is to be
called "exploding soliton era",\cite{TDLee} to be described in the
next section (Sect. 4). The supercooling temperature $T_s$,
presumably $\sim 0.95 T_c$ (to be determined more reliably in the
future), refers to the situation where the bubbles begin to
saturate. We call it the "colliding soliton era". This is to be
described in Section 5. Presumably it would be over when $T\leq
T_s - (T_c-T_s)$ or longer. So, the cosmological QCD phase
transition would be over when the Universe was $10^{-4}sec$ old.
The important things is that, because the phase transition is of
the first order, it releases a huge amount of energy:

\begin{equation}
\rho_{vac}=1.0163\times 10^{14}gm/cm^3,
\end{equation}
clearly much bigger than the radiation density (cf. Eq. (14)).
This quantity is in fact the same as "the zero-point energy". {\bf
That is why we question the radiation-dominated universe.}

When the low-temperature bubbles fill up the space, the
neighboring two bubbles would in general be labelled by different
$\theta_{i,j}$ representing different but degenerate vacua - we
assume that there are infinite many choices of $\theta$; they are
degenerate but complete equivalent. The domain wall is used to
separate the two regions. Three different regions would meet in a
line - which we call a vortex. We have to estimate the total
energy associated with the domain walls and the vortices -
particularly when these objects persist to live on for a "long"
time - say, $\tau \gg 10^{-4} sec$. These domain walls and
vortices are governed, in the QCD phase transition in the early
Universe, by the QCD dynamics - this is an important point; if
not, what else? It is a tough question to figure out how long the
Universe would stabilize itself through QCD dynamics and others;
my rough guess is from a few seconds to years, say $\tau$.

For the moment, QCD enables us to make some estimates. Let us
focus on $t \sim 10^{-4}sec$, where $\rho_m=1.0 \times 10^2\,
gm/cm^3$. Or, considering a unit volume of $1.0\, cm^3$, the
amount of the matter would be $100 gm$ or $5.609\times
10^{31}GeV/c^2$. One proton or neutron weighs about $1\, GeV/c^2$
so, in a volume $1.0\, cm^3$ at $t\sim 10^{-4}sec$, we had at
least $5.609\times 10^{31}$ baryons or, in the MIT bag model
language, $5.609 \times 10^{31}$ bags or $R=1.0\, fermi$ false
vacua associated with the system. To begin with, all the excited
baryons and mesons, including topological objects, and
multi-baryons, all have equal opportunities. But, remembering $1\,
cm^3 = 10^{39} fermi^3$, most space had to collapse into the true
vacua with different $\theta_i$.

\section{Exploding Solitons}

We begin our study by examining an isolated bubble - expanding,
that is, inside the bubble, it is the true vacuum labelled by some
$\theta$; outside the bubble, the false vacuum; we are thinking of
the Universe cooling down and expand. That is, how the bubble
nucleates in the false vacuum (high-temperature). Remember that
this happened in the period of time when $T_c\sim T \leq T_s$.

Consider a spherical wall of radius $R$ and thickness $\Delta$
separating the true vacuum inside from the false vacuum outside.
The energy density difference of the vacua is $B$, the bag
constant in the most simplified situation, and the energy $\tau$
per unit area associated with the surface tension on the
separating wall is a quantity to be calculated but nevertheless is
small compared to the latent heat. If the wall expands outward for
a distance $\delta R$, then the energy budget arising from the
vacuum change is
\begin{equation}
B\cdot 4\pi R^2 \cdot \delta R - \tau\cdot 4\pi \{(R+\delta R)^2 -
R^2\} = -p \delta V,
\end{equation}
where $p$ is the pressure and is so defined that a negative
pressure would push the wall outward. (We use the notation $\tau$
here, since $\sigma$ and $\rho$ are reserved for other purposes.)

When the surface tension energy required for making the wall
bigger is much less than the latent heat required from the
expansion of the bubble, the bubble of the stable vacuum inside
will grow in an accelerating way, possibly resulting in explosive
growth of the bubble. The scenario may be as follows: When the
universe expands and cools, to a temperature slightly above the
critical temperature $T_c$, bubbles of lower vacua will nucleate
at the spots where either the temperature is lower, and lower than
$T_c$, or the density is higher, and higher than the critical
density $\rho_c$. As the universe continues to expand and cool
further, most places in the universe have the temperature slightly
below $T_c$; that is, the destiny arising from eternal expansion
of the universe is driving the average temperature of the entire
universe toward below the critical temperature. The universe must
find a way to convert itself entirely into another vacuum, the
true vacuum at the lower temperature.

Therefore, we have a situation in which bubbles of true vacua pop
up (nucleate) here and there, now and then, and each of them may
grow explosively in the environment made of the false vacuum for
now, but previously the true vacuum when the temperature was still
near the critical temperature $T_c$. In the expanding universe
which cooled down relatively rapidly, i.e. from $T_c$ to the
supercooling temperature $T_s$, the situation is awfully
complicated. When the temperature becomes lower than $T_s$, the
problem can be modelled, in the simplest way, by characterizing
the vacuum structure by a complex scalar field interacting via the
potential $V(\phi)$:
\begin{equation}
V(\phi)= {\mu^2\over 2} \phi^*\phi + {\lambda\over 4}
(\phi^*\phi)^2,\qquad \mu^2 <0,\quad \lambda >0.
\end{equation}
For $T>T_c$, we have $\mu^2(T)>0$ and $\lambda>0$, so it is
between $T_c$ and $T_s$ when the situations are awfully
complicated (and we try to avoid in this paper). Note also that,
in the complex scalar field description, the true vacua have
degeneracy described by a continuous real parameter $\theta$.
$\phi =0$ everywhere in the spacetime describes the false vacuum
for the universe at a temperature below the critical temperature
$T_c$. Consider the solution for a bubble of true vacuum in this
environment. It is required that the field $\phi$ must satisfy the
field equation everywhere in spacetime, including crossing the
wall of thickness $\Delta$ to connect smoothly the true vacuum
inside and the false vacuum outside. This is why we may call the
bubble solution "a soliton", in the sense of a nontopological
soliton of T.D. Lee's. However, the soliton grows in an
accelerating way, or the name "exploding soliton".

The situation must have changed so explosively that at a very
short instant later the universe expands even further and cools to
even a little more farther away from $T_c$ and most places in the
universe must be in the true vacuum, making the previously false
vacuum shrink and fractured into small regions of false vacua,
presumably dominantly in spherical shape, which is shrinking in an
accelerating way, or "implosively". Using again the complex scalar
field as our language, we then have "imploding solitons".

In what follows, we attempt to solve the problem of an exploding
soliton, assuming that the values of both the potential parameters
$\mu^2$ and $\lambda$ are fairly stable during the period of the
soliton expansion. The scalar field must satisfy:
\begin{equation}
{1\over r^2}{\partial\over\partial r} \big( r^2 {\partial\phi\over
\partial r}\big)- {\partial^2\phi\over \partial t^2} =
V^\prime(\phi).
\end{equation}
The radius of the soliton is $R(t)$ while the thickness of the
wall is $\Delta$:
\begin{eqnarray}
\phi &=& \phi_0, \quad for\quad r <R_0+vt -{\Delta\over 2},\nonumber \\
     &=& 0, \qquad for\quad r> R_0+vt+{\Delta\over 2},
\end{eqnarray}
with $R(t)=R_0 +vt$ and $v$ the radial expansion velocity of the
soliton.

We may write
\begin{equation}
\phi \equiv f(r+vt);\qquad w\equiv (1-v^2)r,
\end{equation}
so that the field equation becomes
\begin{equation}
{d^2f\over dw^2} + {2\over w} {df\over dw} = (1-v^2)^{-1} \lambda
f(\mid f\mid^2 - \phi_0^2).
\end{equation}
We will be looking for a solution of $f$ across the wall so that
it connects smoothly the true-vacuum solution inside and the false
vacuum solution outside.

Introducing $g \equiv w f(w)$, we find
\begin{equation}
g^{\prime\prime}=(1-v^2)^{-1} \lambda g\{ \mid{g\over w}\mid^2
-\phi_0^2\},
\end{equation}
an equation which we may solve in exactly the same manner as the
colliding-wall problem to be elucidated in the next section.

\section{Colliding Walls - Formation of "Pasted" Domain Walls}

When bubbles of true vacua grow explosively, the nearby pair of
bubbles will soon squeeze or collide with each other, resulting in
merging of the two bubbles while producing cosmological objects
that have specific coupling to the system. The situation is again
extremely complicated. Remember that this happened when $T\sim
T_s$, not too long after.

We try to disentangle the complexities by looking at between the
two bubble walls that are almost ready to touch and for the
initial attempt neglecting the coupling of the vacuum dynamics to
the matter content. Between the two bubble walls, especially
between the centers of the two bubbles, it looks like a problem of
plane walls in collision - and this is where we try to solve the
problem to begin with.

In fact, we have to consider one bubble first - the spherical
situation as in the previous section but the bubble is "very"
large we could look at the $z$-direction in the sufficiently good
plane approximation (i.e. all bubble surfaces are just like
planes). At this point, we have one wall, with thickness $\Delta$,
moving with velocity $v$ in the $z$ direction; on the left of the
wall is the false vacuum, and on the right the true vacuum.

The wall, of thickness $\Delta$, separates the true vacuum on one
side from the false vacuum on the other side of the wall. For the
sake of simplicity, the wall is assumed parallel to the
$(xy)-$plane and are infinite in both the $x$ and $y$ directions.
In addition, at some instant the wall is defined between
$z=-{\Delta \over 2}$ and $z={\Delta \over 2}$ with the
instantaneous velocity $+v$.

For $z > R + {\Delta \over 2}$ and all $x$ and $y$, the complex
scalar field $\phi$ assumes $\phi_0$, a value of the true vacuum
(the ground state). On the other hand, for $z < -R - {\Delta \over
2}$ and all $x$ and $y$, the complex scalar field $\phi$ assumes
$\phi=0$, the false vacuum. As indicated earlier, the field $\phi$
must satisfy the field equation everywhere in spacetime:
\begin{equation}
{\partial^2\phi\over \partial z^2}- {\partial^2\phi\over\partial
t^2} = V^\prime(\phi).
\end{equation}

We may write the wall on the right hand side but moving toward the
left with the velocity $v$:
\begin{equation}
\phi = f(z-vt), \qquad for\,\, z-vt > 0,\,\, t<R/v.
\end{equation}
so that
\begin{equation}
(1-v^2)f^{\prime\prime}=\lambda f(\mid f\mid^2 -\sigma^2), \qquad
\sigma\equiv \mid \phi_0\mid >0.
\end{equation}

In fact, we are interested in the situation that the function in
Eq. (20) is complex:
\begin{equation}
f\equiv u e^{i\theta},
\end{equation}
so that, with $\tilde\lambda \equiv \lambda/(1-v^2)$,
\begin{equation}
u^{\prime\prime} -u (\theta')^2 = {\tilde\lambda}u(u^2 -\sigma^2),
\end{equation}
\begin{equation}
2u'\theta' + u\theta^{\prime\prime}=0.
\end{equation}
Integrating the second equation, we find
\begin{equation}
u^2 \theta' = K,
\end{equation}
with $K$ an integration constant. The equation for $u$ is thus
given by
\begin{equation}
u^{\prime\prime}={K\over u^3}+{\tilde\lambda}u(u^2-\sigma^2),
\end{equation}
provided that the $\theta$ function is defined (in the region of
the true vacuum and the wall).

Let us try to focus on the last two basic equations - for $u$ and
$\theta$, say, as the functions of $\xi$ (e.g. $\xi = z \pm vt$).
For $\xi \ge \Delta$, we have $\phi = \sigma e^{i\theta}$ (the
true vacuum) and, for $\xi < 0$, we have $\phi = 0$ (the false
vacuum; with $\theta$ undetermined). We find, for $\xi \rightarrow
0^+$,
\begin{equation}
\theta = {1\over 2} \sqrt {-K} (ln \xi)(1+F(\xi))+C_0,
\end{equation}
with $C_0$ a constant and $F(\xi)$ regular near $\xi \sim 0$.
Therefore the $\theta(\xi)$ function could be "mildly singular" or
blow up near $\xi \sim 0$ - this is in fact a very important
point.

Of course, the equation for $u$ can be integrated out to obtain
the result. For the "wall" region (i.e. $0<\xi<\Delta$), the
solution reads as follows:
\begin{equation}
\xi={\sigma^2\over 2}\int_0^{u^2/\sigma^2}{dy\over \sqrt{-K+\alpha
y-2\beta y^2 +\beta y^3}},
\end{equation}
with
\begin{equation}
\Delta={\sigma^2\over 2}\int_0^1{dy\over \sqrt{-K+\alpha y-2\beta
y^2 +\beta y^3}}.
\end{equation}
Here $\beta\equiv{\tilde\lambda\over 2}\sigma^6$, and $K$ and
$\alpha$ parameters related to the integration constants. Of
course, the solution in true-vacuum region can be obtained by
extension.

In the wall region, we could compute the surface energy per unit
area (i.e. surface tension mentioned earlier in Eq. (16)):
\begin{equation}
\tau = \int_0^\Delta d\xi {1\over 2}\{ (u')^2 + u^2
(\theta^\prime)^2\},
\end{equation}
some integral easy to calculate.

There is an important note - that is, the solution for $\phi$
obtained so far applies for the true vacuum and the wall, and
which is continuous in the region; how about the false vacuum?
This is an important question because in the false vacuum we know
that $u = 0$ but $\theta$ is left undetermined. So, in first-order
phase transitions we have certain function undefined in the
false-vacuum region(s). This is a crucial point to keep in mind
with.

As a parenthetical footnote, we note that the equation for the
exploding or imploding spherical soliton, Eq. (22), may be
integrated and solved in an identical manner.

Now let us focus on the merge of the two bubbles - the growing of
the two true-vacuum bubbles such that the false-vacuum region gets
squeezed away. This is another difficult dynamical question. In
fact, we can make the false-vacuum region approaching to zero,
i.e., the region with the solution $u=0$ gets squeezed away; one
true-vacuum region with $\theta_1$ and $\Delta_1$ (the latter for
the wall) is connected with the one with $\theta_2$ and $\Delta_2$
- we could use $(K_1,K_2)$ to label the new boundary; to be
precise, we could call it "the pasted domain wall" or "the patched
domain wall". It is in fact two walls pasted together - if we look
at the boundary condition in between, we realize that the
structure would persist there for a while to go. The pasted domain
wall could evolve further but this may not be relevant for
counting the energies involved. The evolved forms of the pasted
domain walls could be determined by the topology involved - for
the purpose of this paper, we can ignore this fine aspect.

Suppose that the cosmological QCD phase transition was just
completed - we have to caution that, not everywhere, the false
vacua be replaced by the true vacua so that in between the walls
be replaced (approximately) by the pasted domain walls. There are
places for color-singlet objects (i.e. hadrons) which quarks and
gluons tried to hide; these places are still called by the "false
vacua" with the volume energies. Thus, the volume energy, i.e. $B$
in Eq. (16) or defined suitably via $\lambda$ and $\mu^2$ (in Eq.
(17)), or at least some portion of it, may convert itself into the
surface energy and others - $B = 57\, MeV/ fm^3$ using the
so-called "bag constant" in the MIT bag model \cite{MIT} or
Columbia bag model \cite{TDLee}.

This energy density $B=57 MeV/fm^3 =1.0163 \times 10^{14}gm/cm^3$
is huge as compared to the radiation density $\rho_\gamma$ (which
is much bigger than the matter density $\rho_m$) at that time, $t
\sim 10^{-5} \sim 10^{-4}sec $ (see Eqs. (13)-(15)). Some exercise
indicates that this quantity of energy is exactly the latent
"heat" or energy released in the first-order phase transition.

The cosmological QCD phase transition should leave its QCD mark
here - since the volume energy that stays with the "false vacuum"
is simply reduced because the volumes with the "false vacua" are
greatly reduced - but not eliminated because quarks and gluons,
those objects with colors, still have some places to go (or, to
hide themselves).

\section{Possible Connection with the Dark Matter}

Let us begin by making a simple estimate - the expansion factor
since the QCD phase transition up to now. The present age of the
Universe is $13.7$ billion years or $13.7\times 10^9 \times 365.25
\times 24 \times 3600$ or $4.323 \times 10^{17}$ seconds. As
indicated earlier (cf. the end of Sec. 2), about the first $10^9
sec$ period of the hot big bang is previously-believed
radiation-dominated. Consider the length $1.0\, fermi$ at $t\sim
10^{-5}sec$, it will be expanded by a factor of $10^7$ up to $t
\sim 10^9 sec$ (radiation-dominated) and expanded further by
another factor of $5.7\times 10^5$ until the present time - so, a
total expansion factor of $5.7\times 10^{12}$; changing a length
of $2\, fermi$ at $t\sim 10^{-5}sec$ into a distance of $1\, cm$
now. A proton presumably of $R=1\, fermi$ at $t\sim 10^{-4} sec$
should be more or less of the same size now; or, the bag constant
or the energy associated with the false vacuum should remain the
same.

What would happen to the pasted or patched domain walls as formed
during the cosmological QCD phase transition? According to Eqs.
(30) and (31) together with Eq. (32), we realize that the
solutions in previously two different true-vacuum regions cannot
be matched naturally - unless the K values match accidently. On
the other hand, it is certain that the system cannot be stretched
or over-stretched by such enormous factor, $10^{12}$ or $10^{13}$.

As we said earlier, at some point after the supercooling
temperature $T_s$, say, at $T_s-\lambda(T_c-T_s)$ (with $\lambda$
an unknown factor, presumably $\lambda\gg 1$), the system (the
Universe) was temporarily stabilized since most of the pasted or
patched domain walls had no where to go. Remember that all these
happened in a matter of a fraction of $10^{-4}sec$, as judging
from the size of $T_c$ and $T_s$. The next thing to happen is
probably the following.

We believe that the field $\phi$, being effective, cannot be
lonely; that is, there are higher-order interactions such as
\begin{equation}
c_0\phi G_\mu^a G^{\mu,a},\quad c_1 \phi GGG,\, ...,\qquad d_0 \phi
{\bar \psi}\psi,
\end{equation}
some maybe being absent because of the nature of $\phi$. In other
words, we may believe that the strong interactions are primarily
responsible for the phase transition in question, such that the
effective field $\phi$ couples to the gluon and quark fields; the
details of the coupling are subject to investigations.

That is, when the field $\phi$ responsible for the pasted or
patched domain walls is {\it effective} - the $\phi$ field
couples, in the higher-order (and thus weaker) sense, to the gluon
and quark fields. It is very difficult to estimate what time is
needed for pasted domain walls to disappear, if there are no
nontrivial topology involved. If there is some sort of nontrivial
topology present, there should left some kind of topological
domain nugget - however, energy conservation should tell us that
it cannot be expanded by too many orders (but our Universe did
expand for many many orders of magnitude). I would guess that it
takes about from a fraction of a second to several years (from the
strong interaction nature of the problem), but certainly before
the last scattering surface (i.e. $3.79\times 10^5$ years).

To summarize, the energy associated with the cosmological QCD
phase transition, mainly the vacuum energy associated with the
false vacuum, disappeared in several ways, viz.: (1) the bag
energies associated with the baryons and all the other
color-singlet objects, (2) the energies with all kinds of
topological domain nuggets or other topological objects, and (3)
the decay products from pasted or patched domain walls with
trivial topology.

Let us begin with the critical temperature $T = T_c \approx 150\,
MeV$ or $t \approx 3.30 \times 10^{-5} sec$. At this moment, we
have
\begin{equation}
\rho_{vac}=1.0163\times 10^{14} gm/cm^3,\quad \rho_\gamma
=5.88\times 10^9 gm/cm^3,\quad \rho_m=6.51\times 10^2 gm/cm^3.
\end{equation}
Here the first term is what we expect the system to release - the
so-called "latent heat"; I call it "latent energy" for obvious
reasons. The identification of the latent "heat" with the bag
constant is well-known in Coulomb bag models \cite{TDLee}.

This can be considered just before the cosmological QCD phase
transition which took place - at the moment the energy components
which we should take into consideration.

As time went on, the Universe expanded and the temperature cooled
further - from the critical temperature to the supercooling
temperature ($T_s \sim 0.95 \times T_c$ with the fraction 0.95 in
fact unknown) and even lower, and then the cosmological QCD phase
transition was complete. When the phase transition was complete,
we should estimate how the energy $\rho_{vac}$ is to be divided.

Let's assume that the QCD phase transition was completed at the
point $T_s$ (in fact maybe a little short after $T_s$). Let's take
$T_s=0.95\,T_c$ for simplicity. We would like to know how the
energy $\rho_{vac}$ is to be divided. First, we can estimate those
remained with the baryons and other color-singlet objects - the
lower limit is given by the estimate on the baryon number density
(noting that one baryon weighs about $1.0 GeV/c^2$):
\begin{equation}
\rho_m=6.51\times 10^2 gm/cm^3\times 0.5609 \times 10^{24}
GeV/c^2/gm=3.65\times 10^{26} GeV/c^2/cm^3.
\end{equation}
So, in the volume $1.0 cm^3$ or $10^{39} fermi^3$, we have at
least $3.65\times 10^{26}$ baryons. One baryon has the volume
energy (i.e. the bag energy or the false vacuum energy) $57
MeV/fermi^3 \times {4\over 3} \pi (1.0 fermi)^3$ (which is $238.8
MeV$). So, in the volume $1.0 cm^3$, we have at least $238.8 MeV
\times 3.65\times 10^{26}$ or $8.72 \times 10^{25} GeV$ in baryon
bag energy. Or, in different units $8.72\times 10^{25} /
(0.5609\times 10^{24})$ $gm/c^2$ or $155.5 gm/c^2$. {\it Only a
tiny fraction of $\rho_{vac}$ is to be hidden in baryons or other
color-singlet objects after the QCD phase transition in the early
Universe}.

So, where did the huge amount of the energy $\rho_{vac}$ go? In
the beginning of the end of the phase transition, the pasted
domain walls with the huge kinetic energies seem to be the main
story. A pasted domain wall is forming by colliding two domain
walls while eliminating the false vacuum in between. The kinetic
energies associated with the previously head-on collision become
vibration, center-of-mass motion, etc. Of course, the pasted
domain walls would evolve much further such as through the
decaying interactions given earlier or forming the "permanent"
structures. In any case, the total energy involved is known
reasonably - a large fraction of $\rho_{vac}$, much larger than
the radiation $\rho_\gamma$ (with $\rho_m$ negligible at this
point).

The story is relatively simple when the cosmological QCD phase
transition was just completed and most "pasted" domain walls still
have no time to evolve. We return to Eqs. (2) and (3) (i.e.
Einstein equations) for the master equations together with the
equation of state with $\rho$ and $p$ determined by the
energy-momentum tensor:
\begin{equation}
T_{\mu\nu}^\phi=g_{\mu\alpha}{\partial {\it L}\over \partial
(\partial_\alpha \phi)} \partial_\nu \phi - {\it L}g_{\mu\nu}.
\end{equation}
Further analysis indicates that the equation of state for the
"pasted" or "patched" domain walls is nothing unusual - the reason
is that we are working in the real four-dimensional space-time and
all of the objects are of finite dimensions in all the directions.
The "domain walls" discussed by us are for real and cannot be
stretched to infinity in a certain dimension.

In fact, there is certain rule which one cannot escape. Let assume
a simple equation of state, $\rho=w p$, for simplicity and come to
look at Eq. (5). Let's consider the situation in which there is no
curvature $k=0$ and the cosmological constant $\lambda$ is not yet
important.
\begin{equation}
2{\ddot R\over R} +(1+3w){{\dot R}^2\over R^2}=0,
\end{equation}
which yields
\begin{equation}
R \propto t^n,
\end{equation}
with $n={2\over 3}\cdot{1\over 1+w}$.

From the equation of continuity, $d(\rho R)+ pd(R^3)=0$, it is
easy to obtain $\rho \propto R^{3(1+w)}$. Thus, we deduce that,
under very general situations, the density behaves like
\begin{equation}
\rho = C t^{-2},
\end{equation}
where the constant $C$ is related to $w$ in the simplified
equation of state. It is clear that the limit to $w=-1$ (the
cosmological constant) is a discontinuity.

Of course, Eq. (4) is still valid:
\begin{equation}
{\ddot R \over R}=-{4\pi G_N\over 3}(\rho + 3p)+{\Lambda\over 3}.
\end{equation}
This has an important consequence - the idea of the previous
universe expansion usually based on the radiation alone from
$t\sim 10^{-10}\, sec$ (after the cosmological electroweak phase
transition had taken place) to $t\sim 10^9 \,sec$ (when it was
close that $\rho_\gamma=\rho_m$) has to be modified because the
latent energy $\rho_{vac}$ was about $2\times 10^5$ times the
radiation energy at the moment of the cosmological QCD phase
transition.

Shown in Fig. 1 is our main result - even though it is a
qualitative figure but it tells us a lot. At $t\sim 3.30\times
10^{-5}\, sec$, where did the latent energy $10^{14}gm/cm^3$
evolve into? We should know that the curve for $\rho_\gamma$, for
massless relativistical particles, is the steepest in slope. The
other curve for $\rho_m$ is the other limit for matter (which
$P\approx 0$). In this way, the latent energy is connected
naturally with the curve for $\rho_{DM}$ - in fact, there seems to
be no other choice. Remember that $\rho \propto t^{-2}$ except the
slope for different types of "matter".

\begin{figure}[h]
\centering
\includegraphics[width=4in]{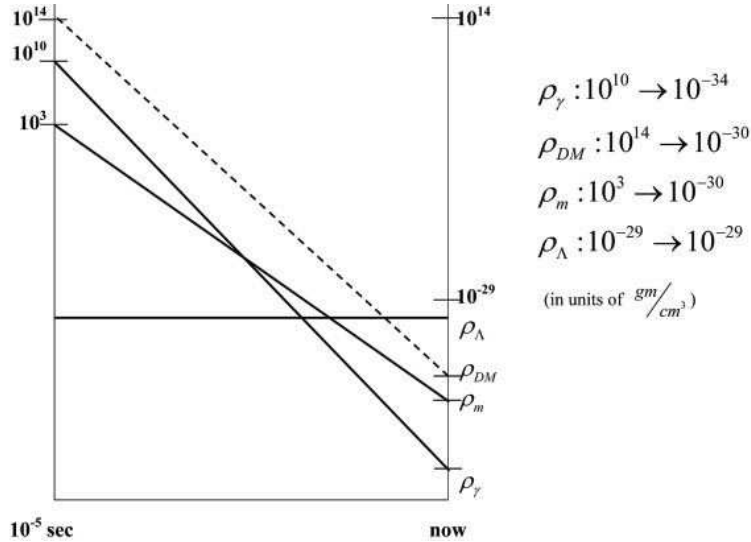}
\caption{The various densities of our universe versus time.}
\end{figure}

Coming back to Eq. (43) or (4), we could assume for simplicity
that when the cosmological QCD just took place the system follows
with the relativistical pace (i.e. $P=\rho/3$) but when the system
over-stretched enough and had evolved long enough it was diluted
enough and became non-relativistic (i.e. $P\approx 0$). It so
happens that in both cases the density to the governing equation,
Eq. (43) or (4), looks like $\rho \propto t^{-2}$ although it is
$R \propto t^{1\over 2}$ followed by $R \propto t^{2\over 3}$.

It is so accidental that what we call "the radiation-dominated
universe" is in fact dominated by the latent energy from the
cosmological QCD phase transition in the form of "pasted" or
"patched" domain walls and the various evolved objects. In our
case, the transition into the "matter-dominated universe", which
happened at a time slightly different from $t\sim 10^9 sec$,
occurred when all the evolutions of the pasted domain walls ceased
or stopped. In other words, it is NOT the transition into the
"matter-dominated universe", as we used to think of.

In fact, the way of thinking of the "dark matter", or the majority
of it, turns out to be very natural. Otherwise, where did the $25
\%$ content of our universe come from? Of course, one could argue
about the large amount of the cosmological QCD phase transition.
We believe that the curves in Fig. 1 make a lot of sense.

Of course, one should ask what would happen before the
cosmological QCD phase transition. It might not be the
radiation-dominated. I believe that it opens up a lot of important
and basic questions.

\section{Summary}

To sum up, we tried to illustrate how to describe the QCD phase
transition in the early Universe, or the cosmological QCD phase
transition.

The scenario that some first-order phase transitions may have
taken place in the early Universe offers us one of the most
intriguing and fascinating questions in cosmology. In fact, the
role played by the latent "heat" or energy released in the process
is highly nontrivial.

In this paper, I take the wisdom that the QCD phase transition,
which happened at a time $t\approx 3.30 \times 10^{-5}\,sec$ or at
the temperature of about $150\, MeV$ and accounts for confinement
of quarks and gluons to within hadrons in the true QCD vacuum,
would be of first order. Thus, it is sufficient to approximate the
true QCD vacuum as one of degenerate $\theta$-vacua and when
necessary we try to model it effectively via a complex scalar
field with spontaneous symmetry breaking. We examine how and how
long "pasted" or "patched" domain walls were formed, how and how
long such walls evolve further, and why the majority of dark
matter might be accounted for in terms of these evolved objects.

Our central result could be summarized by Fig. 1 together with the
explanations. Mainly, we are afraid that the "radiation-dominated"
epoch and the "matter-dominated" epoch, in the conventional sense,
could not exist once the cosmological QCD phase transition took
place. That also explains why there is the $25\%$ dark-matter
content, larger than the baryon content, in our present universe.

{\it Footnote: During the period which the paper is revised and
refereed, some early version of this paper has been accepted for
published in Modern Physics Letters A.}

\section*{Acknowledgments}
The Taiwan CosPA project is funded by the Ministry of Education
(89-N-FA01-1-0 up to 89-N-FA01-1-5) and the National Science
Council (NSC 95-2752-M-002-007-PAE). This research is also
supported in part as another National Science Council project (NSC
95-2119-M-002-034).

\end{document}